\documentclass[english]{article}
\usepackage[T1]{fontenc}
\usepackage{textcomp}
\usepackage[latin9]{inputenc}
\usepackage{amsmath}
\usepackage{amssymb}

\makeatletter

\newcommand{\noun}[1]{\textsc{#1}}

\newcommand{\lyxaddress}[1]{
	\par {\raggedright #1
	\vspace{1.4em}
	\noindent\par}
}

\makeatother

\usepackage{babel}
\begin{document}
\title{Signature of matter-field coupling in quantum-mechanical statistics}
\author{Ana María Cetto$^{*}$ and Luis de la Peña}
\maketitle

\lyxaddress{Instituto de Física, Universidad Autónoma de México, Mexico City,
Mexico.}

$^{*}$Corresponding author: ana@fisica.unam.mx
\begin{abstract}
The connection between the intrinsic angular momentum (spin) of particles
and the quantum statistics is established by considering the response
of identical particles to a common background radiation field. For
this purpose, the Hamiltonian analysis previously performed in stochastic
electrodynamics to derive the quantum description of a one-particle
system is extended to a system of two identical bound particles subject
to the same field. Depending on the relative phase of the response
of the particles to a common field mode, two types of particles are
distinguished by their symmetry or antisymmetry with respect to particle
exchange. While any number of identical particles responding in phase
can occupy the same energy state, there can only be two particles
responding in antiphase. Calculation of bipartite correlations between
the response functions reveals maximum entanglement as a consequence
of the parallel response of the particles to the common field. The
introduction of an internal rotation parameter leads to a direct link
between spin and statistics and to a physical rationale for the Pauli
exclusion principle.

\end{abstract}

\section{Introduction}

The statistics of identical particles is one of the most fundamental
quantum features; all quantum particles are known to obey either Fermi-Dirac
or Bose-Einstein statistics. It is also well known that the intrinsic
angular momentum (spin) of a particle determines its statistics, and
vice versa, with integral-spin particles being bosons and half-integral-spin
particles being fermions. The symmetrization postulate and the spin-statistics
theorem are central to a number of key quantum applications, including
the whole of atomic, molecular and nuclear physics, and quantum statistical
physics. And yet, a century after their establishment,\cite{Pauli25}-\cite{Dirac26}
they continue to be taken as empirical facts, mathematically justified.
All experimental data known are consistent with Pauli's exclusion
principle, and experiments continue to be carried out to find possible
violations of it.\cite{Kaplan20} Pauli himself, who gave the first
formal proof of the spin-stastistics theorem in 1925, expressed his
dissatisfaction with this state of affairs two decades later;\cite{Pauli46,Pauli50}
but explanations continue to rely mainly on formal arguments based
on topological properties, group-theoretical considerations and the
like. 

All this leads to the conclusion that the physical underpinning of
quantum statistics remains to be elucidated. What makes the state
vectors of identical multipartite systems be either symmetric or antisymmetric?
What is the mechanism that ``binds'' identical particles in such
a way that they obey either Fermi or Bose statistics? 

The aim of this paper is to provide an answer to these questions based
on general principles and previous results from stochastic electrodynamics
(\noun{sed}). Recent work has shown that considering the interaction
of particles with the electromagnetic radiation field is key to understanding
their quantum behavior.\cite{TEQ} On the one hand, the ground state
of the radiation field ---i.e. the zero point field (\noun{zpf})---
has been identified as the source of quantum fluctuations and as a
key factor in driving a bound system to a stationary state. Second,
the quantum operator formalism has been obtained as the algebra describing
the response of the particle's dynamical variables to the background
field modes responsible for the transitions between stationary states.\cite{FOOP24}
In addition, bipartite entanglement was derived as a consequence of
the interaction of two identical particles with the same field modes.\cite{TEQ}
Against this background, the theory is able to provide us with a physically
grounded explanation of the origin of the symmetry properties of identical
quantum particle systems and the resulting statistics. 

The paper is structured as follows. Section 2 contains a summary of
the \noun{sed} Hamiltonian derivation of the quantum operator formalism,
which gives sense to this formalism as an algebraic description of
the linear (dipolar) resonant response of the particle to a well-defined
set of modes of the background radiation field. In Section 3, the
expression of the dynamical variables of the particle in terms of
linear response coefficients is applied to the analysis of a system
of two identical particles in a stationary state. In Section 4, two
types of particles are identified according to the relative phase
of their coupling to a common field mode in the bipartite case, and
the multipartite case is briefly discussed. In Section 5, it is shown
that the analysis of two-particle correlations leads to entangled
symmetric or antisymmetric state vectors. In Section 6, the intrinsic
rotation is introduced in order to establish the connection between
the spin and the quantum statistics as reflected in the symmetry of
the state vector, leading to the Pauli exclusion principle for particles
with half-integer spin.

\section{Quantum operators as linear response functions\label{QO}}

As shown in \noun{sed,}\cite{TEQ} the dynamics of an otherwise classical,
charged particle immersed in the zeropoint radiation field of energy
$\hbar\omega/2$ per mode (\noun{zpf}) and subject to a binding force
and its own radiation reaction, evolves irreversibly into the quantum
regime, characterized by the stationary states reached as a result
of the average energy balance between radiation reaction and the action
of the background field. In \cite{FOOP24} it was shown by means of
a Hamiltonian analysis of the particle-field system, that the \emph{nature}
of the particle dynamical variables---i.e. the kinematics---changes
in the transition to the quantum regime. In this regime, $x(t),p(t)$
no longer refer to trajectories, but to the linear, resonant \emph{response}
of the particle to the driving force of the background field, which
effects the transitions between stationary states. The radiative transitions
between two states $(n,k)$ involve precisely those field modes to
which the particle responds resonantly. Thus from the initially infinite,
continuous set of canonical field variables ($\mathrm{q},\mathrm{p}$),
only those $(\mathrm{q}_{nk},\mathrm{p}_{nk})$ so defined are relevant
for the description in the quantum regime. Since the memory of the
initial particle variables $x(0),p(0)$ is lost and the dynamics is
now controlled by the field, the Poisson bracket of the particle canonical
variables, which initially is taken with respect to the complete set
of (particle+field) variables, reduces to the Poisson bracket with
respect to the (relevant) field variables, and therefore, for the
particle in a stationary state $n$ (note that roman letters are used
for the canonical field variables),
\begin{equation}
\left\{ x_{n}(t),p_{n}(t)\right\} _{\mathrm{qp}}=1,\label{16}
\end{equation}
where
\[
\left\{ x_{n}(t),p_{n}(t)\right\} _{\mathrm{qp}}=\sum_{k\neq n}\left(\frac{\partial x_{n}}{\partial\mathrm{q}_{nk}}\frac{\partial p_{n}}{\partial\mathrm{p}_{nk}}-\frac{\partial p_{n}}{\partial\mathrm{q}_{nk}}\frac{\partial x_{n}}{\partial\mathrm{p}_{nk}}\right).
\]
Instead of the canonical field variables (the quadratures) $(\mathrm{q}_{nk},\mathrm{p}_{nk})$
it is convenient to use the (dimensionless) normal variables $a_{nk}=\exp(i\phi_{nk})$,
where $\phi_{nk}$ is a random phase, which are related to the former
by
\begin{equation}
\mathrm{q}_{nk}=\sqrt{\frac{\hbar}{2\left|\omega_{kn}\right|}}(a_{nk}+a_{nk}^{*}),\:\ \mathrm{p}_{nk}=-i\sqrt{\frac{\hbar\left|\omega_{kn}\right|}{2}}(a_{nk}-a_{nk}^{*}).\label{14}
\end{equation}
This transformation, which takes into account that the energy of the
field mode of frequency $\omega_{kn}$ is equal to $\hbar\omega_{kn}$,
is the entry point of Planck's constant in the equations that follow.

With the transformation (\ref{14}), the Poisson bracket with respect
to the normal variables becomes
\[
\left\{ x(t),p(t)\right\} _{nn}\equiv\sum_{k\neq n}\left(\frac{\partial x_{n}}{\partial a_{nk}}\frac{\partial p_{n}}{\partial a_{nk}^{*}}-\frac{\partial p_{n}}{\partial a_{nk}}\frac{\partial x_{n}}{\partial a_{nk}^{*}}\right)
\]
 
\begin{equation}
=i\hbar\sum_{k\neq n}\left(\frac{\partial x_{n}}{\partial\mathrm{q}_{nk}}\frac{\partial p_{n}}{\partial\mathrm{p}_{nk}}-\frac{\partial p_{n}}{\partial\mathrm{q}_{nk}}\frac{\partial x_{n}}{\partial\mathrm{p}_{nk}}\right),\label{18}
\end{equation}
and therefore, according to Eq. (\ref{16}), the transformed Poisson
bracket must satisfy
\begin{equation}
\left\{ x(t),p(t)\right\} _{nn}=i\hbar.\label{19-1}
\end{equation}

From this and Eq. (\ref{18}) it is clear that $x_{n}(t),p_{n}(t)$
must indeed be linear functions of the normal variables $\left\{ a_{nk}\right\} ,\:k\neq n$.
Thus, $x_{n}(t)$ becomes expressed in the form (in one dimension,
for simplicity) 
\begin{equation}
x_{n}(t)=x_{nn}+\sum_{k\neq n}x_{nk}a_{nk}e^{-i\omega_{kn}t}\mathrm{+c.c.,}\label{eq:12}
\end{equation}
where the index $k$ denotes any other state that can be reached by
means of a transition from $n$ (hence $k\neq n$) and $\omega_{kn}$
is the corresponding transition frequency. The coefficient $x_{nk}$
is the response amplitude of the particle to the field mode of frequency
$\omega_{kn}$. More generally, since the field variables connecting
different states $n,n'$ are independent random variables, $(\partial a_{nk}/\partial a_{n'k})=\delta_{nn'\text{}}$
(for equal times one may omit the time dependence in the expression),
\begin{equation}
\left\{ x,p\right\} _{nn'}=i\hbar\delta_{nn\text{\textasciiacute}}.\label{20}
\end{equation}
Using Eq. (\ref{eq:12}) for $x_{n}(t)$ and 
\begin{equation}
p_{n}(t)=m\dot{x}_{n}(t)=-im\sum_{k\neq n}\omega_{kn}x_{nk}a_{nk}e^{-i\omega_{kn}t}\mathrm{+c.c.}\label{21'-1}
\end{equation}
to calculate the derivatives involved in Eq. (\ref{18}), one obtains
\begin{equation}
\left\{ x(t),p(t)\right\} _{nn}=2i{\displaystyle m\sum_{k\neq n}}\omega_{kn}\left|x_{nk}\right|^{2}=i\hbar.\label{19}
\end{equation}
For $x$ and $p$ real, $x_{nk}^{*}(\omega_{nk})=x_{kn}(\omega_{kn}),\ p_{nk}^{*}(\omega_{nk})=p_{kn}(\omega_{kn}),\ a_{nk}^{*}(\omega_{nk})=a_{kn}(\omega_{kn})$.
This allows us to write Eq. (\ref{20}) in the explicit form
\begin{equation}
{\displaystyle \sum_{k\neq n}}\left(x_{nk}p_{kn'}-p_{n'k}x_{kn}\right)=i\hbar\delta_{nn\text{\textasciiacute}},\label{21}
\end{equation}
and to identify the response coefficients $x_{nk},p_{n'k}$ as the
elements of matrices $\hat{x},\hat{p}$ such that 
\begin{equation}
\left[\hat{x},\hat{p}\right]=i\hbar.\label{22}
\end{equation}
This central result of \noun{sed} reveals the quantum commutator as
\emph{the matrix expression of the Poisson bracket} of the particle
variables ($x_{n},p_{n}$) in any state $n$ with respect to the (relevant)
normal field variables corresponding to the modes $\left\{ nk\right\} $
to which the particle responds resonantly from that state. Further,
Eq. (\ref{19}) is identified with the Thomas-Reiche-Kuhn sum rule,
\begin{equation}
2i{\displaystyle m\sum_{k\neq n}}\omega_{kn}\left|x_{nk}\right|^{2}=i\hbar.\label{23-1}
\end{equation}

In summary, this is the physical essence of the quantum operators:
they describe the linear, resonant response of the (bound) particle
to a well-defined set of field modes. The response coefficients $x_{nk}$
and the transition frequencies $\omega_{kn}$ contained in (\ref{eq:12})
are characteristic of the mechanical system; the corresponding random
normal variables $a_{nk}$ in turn contain the information about the
(stationary, random) background field. By taking the derivatives of
$x_{n}$ and $p_{n}$ given by (\ref{eq:12}) and (\ref{21'-1}) with
respect to $a_{nk},a_{nk}^{*}$ to calculate the Poisson bracket,
the latter are removed from the description; the problem seems to
be reduced to a purely mechanical one, although it is in essence an
electrodynamical one. Once the operator formalism is adopted, the
factor $\hbar$, coming from the transformation expressed in Eq. (\ref{14}),
remains as the only conspicuous imprint left by the field. 

We further note that the structure of the commutator is a direct consequence
of the symplectic structure of the problem; this is a feature of the
Hamiltonian dynamics that remains intact in the evolution from the
initial classical to the quantum regime. The correspondence between
classical Poisson brackets and quantum commutators, insightfully established
by Dirac on formal grounds, thus finds a physical explanation.

To connect with quantum formalism in the Heisenberg representation,
we consider an appropriate Hilbert space on which the operators act.
In the present case, the natural choice is the Hilbert space spanned
by the set of orthonormal vectors $\left\{ \left|n\right\rangle \right\} $
representing the stationary states with energy $\mathcal{E}_{n}$.
With the components of $\hat{x}(t)$ given by $x_{nk}e^{-i\omega_{kn}t}$
(see Eq. (\ref{eq:12})) we have
\begin{equation}
\hat{x}(t)=\sum_{n,k}x_{nk}e^{-i\omega_{kn}t}\left|n\right\rangle \left\langle k\right|.\label{23}
\end{equation}
The matrix elements of $\hat{x}(t)$ are 
\begin{equation}
x_{nk}(t)=\left\langle n\right|\hat{x}(t)\left|k\right\rangle \label{24}
\end{equation}
in the Heisenberg picture, or 
\begin{equation}
x_{nk}(t)=\left\langle n(t)\right|\hat{x}\left|k(t)\right\rangle \label{25a}
\end{equation}
in the Schrödinger picture, where the time dependence has been transferred
to the state vector,
\begin{equation}
\left|n(t)\right\rangle =e^{-i\mathcal{E}_{n}t/\hbar}\left|n\right\rangle .\label{25b}
\end{equation}
 Finally, with the evolution of $x,p$ into operators, the initial
Hamilton equations evolve in the quantum regime into the Heisenberg
equations,
\begin{equation}
\frac{1}{i\hbar}\left[\hat{x},\hat{H}\right]=\hat{\dot{x}},\:\;\frac{1}{i\hbar}\left[\hat{p},\hat{H}\right]=\hat{\dot{p}},\label{26}
\end{equation}
with $\hat{H}=\frac{\hat{p}^{2}}{2m}+\hat{V}$, $\hat{\dot{x}}=\hat{p}/m$
and $\hat{\dot{p}}=-\widehat{(dV/dx)}.$ By taking the matrix element
($nk$) of the first of these equations we confirm that $\omega_{kn}=\left(\mathcal{E}_{n}-\mathcal{E}_{k}\right)/\hbar$,
i.e. that the energy $\hbar\omega_{kn}$ transferred to (or from)
the field to the particle in a transition is equal to the energy difference
between the two stationary states. 

\section{Response of a bipartite system to the background field }

Now consider a system consisting of two identical particles. When
the particles are isolated from each other, they are subject to different
realizations of the background field, in which case their behavior
can be studied separately for each particle, using the procedure above.
However, if they are part of one and the same system, they are subject
to the same realization of the field and, being identical, they respond
to the same set of relevant field modes, whether or not they interact
with each other. In the following we assume that the particles do
not interact directly with each other.

Our purpose is to describe the response of the composite system to
the background field when in a stationary state characterized by the
total energy $\mathcal{E}_{(nm)}=\mathcal{E}_{n}+\mathcal{E}_{m}$
with $\mathcal{E}_{n}\mathcal{\mathrm{\neq}E}_{m}$, the subindices
$n$ and $m$ referring to single-particle states. If particle 1 is
in state $n$ it responds to the set of modes $\left\{ nk\right\} $
and similarly particle 2 in state $m$ responds to the set $\left\{ ml\right\} $,
\begin{equation}
x_{1n}(t)=\sum_{k}e^{i\theta_{nk}^{1}}x_{1nk}a_{nk}e^{-i\omega_{kn}t}\mathrm{+c.c.,\;}\;x_{2m}(t)=\sum_{l}e^{i\theta_{ml}^{2}}x_{2ml\text{}}a_{ml}e^{-i\omega_{lm}t}\mathrm{+c.c.}\label{eq:52}
\end{equation}
where we have added the factor $\exp(i\theta)$ to each term to allow
for the (random) phase of the response of the particle to the field
modes. 

When $n\neq m$, the sums in Eqs. (\ref{eq:52}) involve different,
mutually independent normal variables $a_{nk}$ and $a_{ml}$ except
when $k=m$ and $l=n$, since $a_{nm}=a_{mn}^{*}$. Therefore, the
Poisson bracket of $x_{1}(t)$ and $x_{2}(t)$, calculated in the
state of the composite system ($nm$), reduces to a single term,
\begin{equation}
\left[x_{1},x_{2}\right]_{(nm)}=\left(\frac{\partial x_{1n}}{\partial a_{nm}}\frac{\partial x_{2m}}{\partial a_{nm}^{*}}-\frac{\partial x_{2m}}{\partial a_{nm}}\frac{\partial x_{1n}}{\partial a_{nm}^{*}}\right)=2i\left|x_{nm}\right|^{2}\sin\theta_{nm}^{12}.\label{eq:54}
\end{equation}
Since the particles are identical, the interchange of the labels $1,2$
should not alter the value of the Poisson bracket, therefore this
equation must be equel to zero. This sets an important restriction
on the possible values of the phase difference. Writing
\begin{equation}
\left|\theta_{nm}^{1}-\theta_{nm}^{2}\right|=\left|\theta_{nm}^{12}\right|\equiv\pi\zeta_{nm}^{12},\label{eq:56}
\end{equation}
we see that $\zeta_{nm}^{12}$ must be an integer so that
\begin{equation}
\left[x_{1},x_{2}\right]_{(nm)}=0\;(n\neq m).\label{eq:58}
\end{equation}
Further, with $p_{2}(t)$ obtained from the second Eq. (\ref{eq:52}),
\[
p_{2m}(t)=-im\sum_{l}e^{i\theta_{ml}^{2}}\omega_{lm}x_{2ml\text{}}a_{ml}e^{-i\omega_{lm}t}\mathrm{+c.c.},
\]
the Poisson bracket of $x_{1}(t)$ and $p_{2}(t)$ calculated for
the same state ($nm$) gives
\begin{equation}
\left[x_{1},p_{2}\right]_{(nm)}=\left(\frac{\partial x_{1n}}{\partial a_{nm}}\frac{\partial p_{2m}}{\partial a_{nm}^{*}}-\frac{\partial p_{2m}}{\partial a_{nm}}\frac{\partial x_{1n}}{\partial a_{nm}^{*}}\right)=2im\omega_{mn}\left|x_{nm}\right|^{2}\cos\theta_{nm}^{12}.\label{60}
\end{equation}
In terms of the parameter $\zeta_{nm}^{12}$ defined in Eq. (\ref{eq:56}),
we have 
\begin{equation}
\cos\theta_{nm}^{12}=(-1)^{\zeta_{nm}^{12}},\;\zeta_{nm}^{12}=0,1,2,....\label{61}
\end{equation}
and therefore, from Eq. (\ref{60}),
\begin{equation}
\left[x_{1},p_{2}\right]_{(nm)}=(-1)^{\zeta_{nm}^{12}}2im\omega_{mn}\left|x_{nm}\right|^{2}.\label{eq:62}
\end{equation}
This result shows that a correlation is established between the response
variables of the two particles to the shared field mode ($nm$), for
$n\neq m$; in other words, the field mode serves as a bridge between
the particles and correlates their responses. It is important to note
that Eq. (\ref{eq:62}) involves only the field mode connecting the
two states with $\mathcal{E}_{n}\mathcal{\mathrm{\neq}E}_{m}$, and
it is different from zero only when these states are connected by
a dipolar transition element, $x_{nm}\neq0$. 

Let us now consider two equal particles in the same energy state,
i. e. $n=m$. In this case the particles share all field modes, so
that the Poisson brackets become, by virtue of Eq. (\ref{61}),
\begin{equation}
\left[x_{1},x_{2}\right]_{(nn)}={\displaystyle \sum_{k}}\left(\frac{\partial x_{1n}}{\partial a_{nk}}\frac{\partial x_{2n}}{\partial a_{nk}^{*}}-\frac{\partial x_{2n}}{\partial a_{nk}}\frac{\partial x_{1n}}{\partial a_{nk}^{*}}\right)=2i{\displaystyle \sum_{k}\sin\theta_{nk}^{12}}\left|x_{nk}\right|^{2}=0,\label{64}
\end{equation}
\[
\left[x_{1},p_{2}\right]_{(nn)}={\displaystyle \sum_{k}}\left(\frac{\partial x_{1n}}{\partial a_{nk}}\frac{\partial p_{2n}}{\partial a_{nk}^{*}}-\frac{\partial p_{2n}}{\partial a_{nk}}\frac{\partial x_{1n}}{\partial a_{nk}^{*}}\right)
\]
\begin{equation}
=2i{\displaystyle m\sum_{k}\omega_{kn}\cos\theta_{nk}^{12}}\left|x_{nk}\right|^{2}=2i{\displaystyle m\sum_{k}}(-1)^{\zeta_{nk}^{12}}\omega_{kn}\left|x_{nk}\right|^{2}.\label{66}
\end{equation}

\section{Two families of particles\label{BF}}

Equation (\ref{eq:62}) indicates that there are two distinct types
of identical particles, depending on whether the phase parameter $\zeta_{nm}^{12}$
given by Eq. (\ref{eq:56}) is an even or odd number. Since this condition
applies to all modes that are shared by the two particles, we can
write, using Eq. (\ref{eq:56}): 
\begin{equation}
\zeta_{nm}^{12}=\zeta^{12}=\left|\zeta^{1}-\zeta^{2}\right|,\label{70}
\end{equation}
so that the two types of particles are characterized by \begin{subequations}
\label{BF1} 
\begin{equation}
\mathrm{Type\:B:\;}\zeta^{12}=0,2,4,\ldots,\label{72}
\end{equation}
\begin{equation}
\mathrm{Type\:F:\;}\zeta^{12}=1,3,5,\ldots.\label{74}
\end{equation}
\end{subequations} Note that for all $\zeta^{12}$ to be even in
the first case, the individual $\zeta^{i}$ must be integers; for
all all $\zeta^{12}$ to be odd in the second case, the individual
$\zeta^{i}$ must be half-integers, i.e. \begin{subequations} \label{BF1-1}
\begin{equation}
\mathrm{Type\:B:\;}\left|\zeta^{i}\right|=0,1,2,\ldots\Upsilon_{B},\label{72-1}
\end{equation}
\begin{equation}
\mathrm{Type\:F:\;}\left|\zeta^{i}\right|=\frac{1}{2},\frac{3}{2},\frac{5}{2},\ldots\Upsilon_{F},\label{74-1}
\end{equation}
\end{subequations} where $\Upsilon_{B}$ and $\Upsilon_{F}$ are
the maximum values of the individual $\zeta^{i}$. This means that
$B$ and $F$ stand actually for two families of particles, whose
members are characterized by the respective value of $\Upsilon$.
Since the $\zeta^{i}$ can be positive or negative, for a given $\Upsilon$
there are $g=2\Upsilon+1$ possible different states of the bipartite
system, according to Eqs. (\ref{BF1-1}).

With these results, Eqs. (\ref{eq:52}) take the form
\[
x_{1n}(t)=e^{i\pi\zeta^{1}}\sum_{k}x_{1nk}a_{nk}e^{-i\omega_{kn}t}\mathrm{+c.c.,}
\]
\begin{equation}
x_{2m}(t)=e^{i\pi\zeta^{2}}\sum_{l}x_{2ml\text{}}a_{ml}e^{-i\omega_{lm}t}\mathrm{+c.c.,}\label{75}
\end{equation}
and (\ref{66}) is reduced to
\begin{equation}
\left[x_{1},p_{2}\right]_{(nn)}=(-1)^{\zeta^{12}}i\hbar.\label{70-1}
\end{equation}

Therefore, comparing with the one-particle commutator $\left[x_{1},p_{1}\right]_{(nn)}=i\hbar$,
we note that in the B case particle 2 responds in the same way as
particle 1. Indeed, according to Eq. (\ref{eq:56}), the response
of the two particles to the shared field modes is \emph{in phase},
and a correlation is established between the particles for any pair
of values $-\Upsilon_{B}\leq\zeta^{1},\zeta^{2}\leq\Upsilon_{B}$.
By contrast, for identical particles of type $F$, according to Eq.
(\ref{BF1-1}) $\zeta^{12}$ must be an odd number, hence $\zeta^{1}\neq\zeta^{2}$
and the response of the two particles to the shared field modes is
\emph{in antiphase}.

\subsection{Extension to three or more particles}

Let us briefly analyze the possible correlations for a system composed
of three or more identical particles, in light of the above results.

Take first the case of three type-B particles. When the total energy
$\mathcal{E}_{(nml)}=\mathcal{E}_{n}+\mathcal{E}_{m}+\mathcal{E}_{l}$
with $\mathcal{E}_{n}\mathcal{\mathrm{\neq}E}_{m}\mathcal{\mathrm{\neq}E}_{l}$,
Eq. (\ref{72}) applies and the three particles are pairwise correlated.
According to Eq. (\ref{70-1}) correlation exists also when $\mathcal{E}_{n}\mathcal{\mathrm{\neq}E}_{m}\mathcal{\mathrm{=}E}_{l}$
or $\mathcal{E}_{n}\mathcal{\mathrm{=}E}_{m}\mathcal{\mathrm{=}E}_{l}$,
because the responses of the three particles to common field modes
are always in phase. Therefore, all three particles may in principle
occupy the same state $n$ and respond coherently. The argument can
of course be extended to four or more particles; consequently, there
may in principle be an arbitrary number $N$ of type-B particles in
the same state and respond coherently to the field modes ---like
a well disciplined troop.

In the type-$\mathrm{F}$ case, we have already concluded that particles
1 and 2 respond in antiphase to a common mode and the same applies
of course to any pair of identical particles. When the total energy
$\mathcal{E}_{(nml)}=\mathcal{E}_{n}+\mathcal{E}_{m}+\mathcal{E}_{l}$
with $\mathcal{E}_{n}\mathcal{\mathrm{\neq}E}_{m}\mathcal{\mathrm{\neq}E}_{l}$,
the three particles are pairwise correlated according to Eq. (\ref{74}).
However, when at least two energy levels coincide, two particles respond
in antiphase to the shared modes, which prevents a third one from
responding in antiphase to the same modes and therefore from being
correlated to the other two. Therefore, contrary to the type-B case
there can be no coherent response of more than two type-$\mathrm{F}$
particles in this case.

\section{Field-induced covariance and entanglement\label{Ent}}

To calculate the effect of the background field on the correlation
of the responses we consider two generic dynamical variables associated
with particles 1 and 2; these can be the variables $x(t)$ and $p(t)$
considered so far, a linear combination of them, or any other variable
of the form given by Eq. (\ref{75}), where $n,m$ are as before two
stationary states of the system, with energies $\mathcal{E}_{n},\mathcal{E}_{m}$,
\begin{equation}
f_{1n}(t)=f_{1nn}+e^{i\pi\zeta^{1}}\sum_{k\neq n}f_{1nk}a_{nk}e^{-i\omega_{kn}t}\mathrm{+c.c.,\;}\label{80a}
\end{equation}
\begin{equation}
g_{2m}(t)=g_{2mm}+e^{i\pi\zeta^{2}}\sum_{l\neq m}g_{2ml\text{}}a_{ml}e^{-i\omega_{lm}t}\mathrm{+c.c.,}\label{80b}
\end{equation}

The time-independent terms in these equations represent in each case
the average value of the function, taken over the distribution of
the normal variables $a_{nk}=\exp(i\phi_{nk})$ where $\phi_{nk}$
is a random phase, as mentioned in Section \ref{QO}, 
\begin{equation}
\overline{f_{1n}(t)}=f_{1nn},\;\overline{g_{2m}(t)}=g_{2mm}.\label{82}
\end{equation}
To calculate the correlation we take the average of the product of
$f_{1}(t)$ and $g_{2}(t)$. When particles 1 and 2 do not form part
of the same system, they respond to independent realizations of the
field modes, and therefore the covariance is given by
\begin{equation}
\varGamma(f_{1n}g_{2m})=\left(\overline{f_{1n}(t)}-f_{1nn}\right)\left(\overline{g_{2m}(t)}-g_{2mm}\right)=0,\label{84}
\end{equation}
which simply confirms that the variables are not correlated. 

However, when the particles form a bipartite system they respond to
the same realization of the field modes. To calculate the covariance
in this case we have to take into account the double degeneracy of
the combined state, $\mathcal{E}=\mathcal{E}_{1n}+\mathcal{E}_{2m}=\mathcal{E}_{1m}+\mathcal{E}_{2n}$.
In order to distinguish between the two configurations, we define
\begin{equation}
\mathcal{E}_{C}=\mathcal{E}_{1n}+\mathcal{E}_{2m},\;\mathcal{E}_{D}=\mathcal{E}_{1m}+\mathcal{E}_{2n}.\label{86}
\end{equation}
Let us consider the first case, $\mathcal{E}_{C}=\mathcal{E}_{1n}+\mathcal{E}_{2m},$
and use Eqs. (\ref{80a}) (\ref{80b}) to calculate the average product
of $f_{1}(t)$ and $g_{2}(t)$, which we call $\overline{fg}^{C}$
(the left factor refers always to particle 1 and the right one refers
to particle 2, so that we omit the indices $1,2$ in the following).
Taking into account that for random independent normal variables,
$\overline{a_{ij}a_{jk}}=\overline{a_{ij}a_{kj}^{*}}=\delta_{ik}$
and hence
\begin{equation}
\overline{a_{nk}a_{ml}}=\delta_{nk}\delta_{ml}+\delta_{nl}\delta_{km},\label{88}
\end{equation}
we get
\begin{equation}
\overline{fg}^{C}=f_{nn}g_{mm}+(-1)^{\zeta}f_{nm}g_{mn}.\label{90}
\end{equation}
Similarly, for the D configuration we get
\begin{equation}
\overline{fg}^{D}=f_{mm}g_{nn}+(-1)^{\zeta}f_{mn}g_{nm}.\label{92}
\end{equation}
Since the two configurations have the same weight, the averages of
$f_{1}(t)$ and $g_{2}(t)$ are
\[
\overline{f}=\frac{1}{2}(f_{nn}+f_{mm}),\;\overline{g}=\frac{1}{2}(g_{nn}+g_{mm}),
\]
and the average of the product of $f_{1}(t)$ and $g_{2}(t)$ is given
by
\[
\overline{fg}=\frac{1}{2}\left(\overline{fg}^{C}+\overline{fg}^{D}\right)
\]
\begin{equation}
=\frac{1}{2}\left[f_{nn}g_{mm}+(-1)^{\zeta}f_{nm}g_{mn}+f_{mm}g_{nn}+(-1)^{\zeta}f_{mn}g_{nm}\right].\label{94}
\end{equation}
The covariance is therefore given by
\[
\varGamma(fg)=\overline{fg}-\overline{f}\overline{g}
\]
\begin{equation}
-\frac{1}{4}(f_{nn}-f_{mm})(g_{nn}-g_{mm})+\frac{1}{2}(-1)^{\zeta}\left[f_{nm}g_{mn}+f_{mn}g_{nm}\right].\label{96}
\end{equation}
In this equation, the two contributions to the covariance are of a
very different nature: the first one is a classical covariance of
$f_{1}$ and $g_{2}$ due to the different average values of these
functions in states $n,m$ under the condition of degeneracy, $\mathcal{E}_{1n}+\mathcal{E}_{2m}=\mathcal{E}_{1m}+\mathcal{E}_{2n}.$
The second term, in turn, has no classical counterpart: it is entirely
due to the joint response of particles 1 and 2 to the shared mode
($nm$) and is therefore a signature of the matter-field interaction.
Evidently both particles must respond to the mode ($nm$) for this
term to be different from zero; if any of the two matrices $\hat{f},\hat{g}$
is diagonal, there is no quantum contribution to $\varGamma(fg)$.

\subsection{Emergence of entanglement}

In quantum formalism, entanglement is reflected in the non-factorizability
of the bipartite state vector. Therefore, in order to show the emergence
of entanglement in the present context, we will translate Eq. (\ref{96})
into the language of the product Hilbert space $\mathrm{H_{1}\otimes H_{2}}$,
where $\mathrm{H}_{1},\mathrm{H_{2}}$ are respectively spanned by
the sets of orthonormal state vectors $\left\{ \left|n\right\rangle \right\} $
of particles 1,2 (see Section (\ref{QO}) for the one-particle case).
In the shorthand notation introduced above, configurations $C,D$
are represented by the product state vectors
\begin{equation}
\left|C\right\rangle =\left|n\right\rangle _{1}\left|m\right\rangle _{2},\;\left|D\right\rangle =\left|m\right\rangle _{1}\left|n\right\rangle _{2}.\label{98}
\end{equation}
In this notation, Eq. (\ref{96}) reads
\[
\varGamma(fg)=-\frac{1}{4}(f_{nn}+f_{mm})(g_{nn}+g_{mm})
\]
\begin{equation}
+\frac{1}{2}\left\langle C+(-1)^{\zeta}D\right|\hat{f}\hat{g}\left|C+(-1)^{\zeta}D\right\rangle .\label{100}
\end{equation}
In writing the second term we have used the fact that $(-1)^{\zeta}=\pm1$
according to Eqs. (\ref{BF1}). Note that the average of $fg$ is
now taken over the (normalized) state vector
\begin{equation}
\left|\varPsi\right\rangle \equiv\frac{1}{\sqrt{2}}\left|C+(-1)^{\zeta}D\right\rangle ,\label{102}
\end{equation}
or in terms of the individual state vectors,
\begin{equation}
\left|\varPsi\right\rangle =\frac{1}{\sqrt{2}}\left[\left|n\right\rangle _{1}\left|m\right\rangle _{2}+(-1)^{\zeta}\left|m\right\rangle _{1}\left|n\right\rangle _{2}\right].\label{103}
\end{equation}
As a result, we get
\begin{equation}
\varGamma(fg)=\left\langle \varPsi\right|\hat{f}\hat{g}\left|\varPsi\right\rangle -\left\langle \varPsi\right|\hat{f}\left|\varPsi\right\rangle \left\langle \varPsi\right|\hat{g}\left|\varPsi\right\rangle ,\label{104}
\end{equation}
which is exactly the quantum covariance of $\hat{f}\hat{g}$ calculated
in the entangled state given by Eq. (\ref{103}). The covariance coincides
with the correlation of $f$ and $g$, since the state vector $\left|\varPsi\right\rangle $
is normalised to unity.

We stress that the above calculation is restricted to the case $n\neq m$;
when $n=m$ there is no field mode correlating the responses of the
two particles, so there is no entanglement. On the other hand, if
there is degeneracy, i.e. $\mathcal{E}_{C}=\mathcal{E}_{D}$, the
two-particle system is necessarily in an entangled state if $f_{nm},g_{mn}$
are different from zero, i.e. if the response variables $f,g$ connect
the single-particle states $n,m$. The origin of the entanglement
is thus traced back to the action of the common relevant field mode
$(nm)$, and the responses of the two particles to this mode are maximally
correlated (anticorrelated) according to Eq. (\ref{96}) with $(-1)^{\zeta}=+1$
($-1$). More generally, entanglement occurs whenever there is degeneracy,
be it in energy or any other variable that defines the state of the
bipartite system, as discussed in the next section.

Equations (\ref{102})-(\ref{104}) were previously obtained in the
context of \noun{sed} by a somewhat laborious procedure using the
Hilbert-space formalism. In contrast to such an abstract procedure,
the present derivation has the advantage of keeping track at every
moment of the physical quantities involved, namely the field mode
variables, the particles' response variables and the phase difference
of the responses. 

From Eq. (\ref{103}) it is clear that the two families of identical
particles identified in Section \ref{BF} are distinguished by their
entangled state vectors. The symmetry or antisymmetry of the state
vector is uniquely linked to the phase difference of the responses
of the two particles to the shared field mode. When the coupling is
in phase (type B particles), the state vector is symmetric with respect
to the exchange of particles; when the relative coupling is out of
phase (type F particles), the state vector is antisymmetric. 

It should be stressed that no direct interaction between the components
of the system is involved in the derivation leading to entangled states;
entanglement arises as a result of their indirect interaction via
the shared field modes, and therefore does not entail a non-local
action. 

\section{The Pauli exclusion principle}

\subsection{Introduction of spin}

Among the various proposals that have been made to justify the spin-statistics
theorem, some that are relevant to this work involve the inclusion
of the internal (spin) coordinates among the parameters affected by
the exchange operation; see e.g. Refs. \cite{Hunter705,Jabs10} and
additional references cited in \cite{Jabs10}. In particular, in \cite{Jabs10}
the spin-statistics connection is derived under the postulates that
the original and the exchange wave functions are simply added, and
that the azimuthal phase angle, which defines the orientation of the
spin part of each single-particle spin component in the plane normal
to the spin-quantization axis, is exchanged along with the other parameters. 

In dipolar transitions, atomic electrons interact with field modes
of circular polarization, a fact that is expressed in the selection
rule $\triangle l=\pm1$ and is increasingly exploited for practical
applications in spin-resolved spectroscopy and magneto-optics, see
e.g. Refs. \cite{Okuda11,De21}. Furthermore, the interaction of the
particle with circular polarized modes of the \noun{zpf}, which are
known to have an intrinsic angular momentum equal to $\hbar/2$,\cite{Sobelman79,Mandel95}
was indeed shown in Ref. \cite{JPCS14} to be responsible for the
origin of the electron spin itself. It is reasonable to assume that
a similar mechanism is responsible for the neutron spin, since the
neutron has a magnetic moment that couples to the radiation field.

Therefore, following Refs. \cite{Jabs10,JPCS16}, in order to include
the spin in the present analysis we add an (internal) rotation angle
$\phi$ to the expression for the dynamical variables. Strictly speaking
the problem becomes a three-dimensional one. However, for simplicity,
we can still use our one-dimensional expressions for the dynamical
variables if we decompose the radiation field into (statistically
independent) modes of circular polarization. So instead of (\ref{80a})
and (\ref{80b}) we write 
\begin{equation}
f_{1n}(t,\phi)=e^{i\pi\zeta^{1}}\sum_{k}f_{1nk}a_{nk}e^{i\gamma_{nk}\phi-i\omega_{kn}t}\mathrm{+c.c.,\;}\label{80a-1}
\end{equation}
\begin{equation}
g_{2m}(t)=e^{i\pi\zeta^{2}}\sum_{l}g_{2ml\text{}}a_{ml}e^{i\gamma_{ml}\phi-i\omega_{lm}t}\mathrm{+c.c.,}\label{80b-1}
\end{equation}
where $\gamma_{nk}\phi$ is the difference of two rotation angles,
\begin{equation}
\text{\ensuremath{\gamma_{nk}\phi}}=(\gamma_{n}-\gamma_{k})\phi,\label{105}
\end{equation}
and $\gamma_{n},\gamma_{k}$ stand for counterclockwise (clockwise)
rotation. If $n,m$ are two stationary states of a system of identical
particles, as before, we get for the partial covariances in configurations
$C$ and $D$ (see Eqs. (\ref{90}) and (\ref{92})),
\begin{equation}
\overline{fg}^{C}=f_{nn}g_{mm}+(-1)^{\zeta}f_{nm}e^{i\gamma_{nm}\phi}g_{mn}e^{i\gamma_{mn}\phi},\label{106}
\end{equation}
\begin{equation}
\overline{fg}^{D}=f_{mm}g_{nn}+(-1)^{\zeta}f_{mn}e^{i\gamma_{mn}\phi}g_{nm}e^{i\gamma_{nm}\phi},\label{108}
\end{equation}
and therefore,
\[
\overline{fg}=\frac{1}{2}\left(\overline{fg}^{C}+\overline{fg}^{D}\right)=\frac{1}{2}\left[f_{nn}g_{mm}+f_{mm}g_{nn}\right]
\]
\begin{equation}
+\frac{1}{2}(-1)^{\zeta}\left[f_{nm}e^{i\gamma_{nm}\phi}g_{mn}e^{i\gamma_{mn}\phi}+f_{mn}e^{i\gamma_{mn}\phi}g_{nm}e^{i\gamma_{nm}\phi}\right].\label{110}
\end{equation}
By translating this result into the language of the product Hilbert
space and using Eq. (\ref{105}) we get after some algebra
\begin{equation}
\varGamma(fg)=\left\langle \varPsi\right|\hat{f}\hat{g}\left|\varPsi\right\rangle -\left\langle \varPsi\right|\hat{f}\left|\varPsi\right\rangle \left\langle \varPsi\right|\hat{g}\left|\varPsi\right\rangle ,\label{112}
\end{equation}
where $\left|\varPsi\right\rangle $ stands now for the complete bipartite
state vector, including the internal rotation components,
\[
\left|\varPsi\right\rangle \equiv\frac{1}{\sqrt{2}}\left|e^{-i\gamma_{n}\phi}e^{-i\gamma_{m}\phi}C+(-1)^{\zeta}e^{-i\gamma_{m}\phi}e^{-i\gamma_{n}\phi}D\right\rangle 
\]
\begin{equation}
=\frac{1}{\sqrt{2}}\left|e^{-i\gamma_{n}\phi}\left|n\right\rangle _{1}e^{-i\gamma_{m}\phi}\left|m\right\rangle _{2}+(-1)^{\zeta}e^{-i\gamma_{m}\phi}\left|m\right\rangle _{1}^{-i\gamma_{n}\phi}\left|n\right\rangle _{2}\right\rangle .\label{114}
\end{equation}
In Eq. (\ref{114}), the first angular factor is always associated
with particle 1 and the second with particle 2. This suggests writing
each individual state vector in the form $e^{-i\gamma\phi}\left|n\right\rangle .$
In the quantum language this implies the introduction of two orthonormal
vectors $\left|\gamma\right\rangle =\left|+\right\rangle ,\left|-\right\rangle $
spanning the two-dimensional Hilbert space, $\left|n\right\rangle \left|\gamma\right\rangle \equiv\left|n\gamma\right\rangle $,
so Eq. (\ref{114}) takes the form
\begin{equation}
\left|\varPsi\right\rangle =\frac{1}{\sqrt{2}}\left[\left|n\gamma_{n}\right\rangle _{1}\left|m\gamma_{m}\right\rangle _{2}+(-1)^{\zeta}\left|m\gamma_{m}\right\rangle _{1}\left|n\gamma_{n}\right\rangle _{2}\right].\label{115}
\end{equation}
Since the parameter $\gamma$ is associated with the internal rotation,
we identify it with the spin of the electron, which means that 
\begin{equation}
\gamma_{n,m}=\pm\frac{1}{2}.\label{116}
\end{equation}

\subsection{The connection between spin and symmetry}

We now examine the symmetry properties of the complete entangled state
function (\ref{114}) under particle exchange. When particles 1 and
2 are exchanged, in addition to switching their positions in three-dimensional
space, their internal angles change: particle 1 rotates to the azimuthal
position of particle 2 and vice versa, with both rotations occurring
in the same direction (clockwise or counterclockwise). Consider a
clockwise rotation. Then, as shown in \cite{Jabs10,JPCS16}, when
$\phi_{2}>\phi_{1}$, $\phi_{1}$ transforms into $\phi_{2}$ and
$\phi_{2}$ transforms into $\phi_{1}+2\pi$, so
\begin{equation}
\phi_{2}-\phi_{1}\rightarrow\phi_{1}-\phi_{2}+2\pi,\label{117}
\end{equation}
and $\left|\varPsi\right\rangle $ given by Eq. (\ref{114}) transforms
into
\[
\left|\varPsi\right\rangle _{1\longleftrightarrow2}=\frac{1}{\sqrt{2}}\left|e^{-i\gamma_{m}(\phi+2\pi)}\left|m\right\rangle _{1}e^{-i\gamma_{n}\phi}\left|n\right\rangle _{2}+(-1)^{\zeta}e^{-i\gamma_{n}(\phi+2\pi)}\left|n\right\rangle _{1}e^{-i\gamma_{m}\phi}\left|m\right\rangle _{2}\right\rangle .
\]
Since $\gamma_{n},\gamma_{m}$ are half-integers, the overall effect
of the particle exchange is to multiply the original state vector
by a factor
\begin{equation}
\left|\varPsi\right\rangle _{1\longleftrightarrow2}=(-1)^{\zeta}(-1)^{2\gamma_{n}}\left|\varPsi\right\rangle .\label{124}
\end{equation}
If instead $\phi_{2}<\phi_{1}$, $\phi_{2}$ transforms into $\phi_{1}$
and $\phi_{1}$ transforms into $\phi_{2}+2\pi$, so that
\begin{equation}
\phi_{2}-\phi_{1}\rightarrow\phi_{1}-\phi_{2}-2\pi,\label{117-1}
\end{equation}
and the transformation of the state vector is again given by Eq. (\ref{124}).
Of course, the same result is obtained if the rotation is anticlockwise.
Since particles 1 and 2 are identical, their exchange should have
no effect on the state vector, which implies that 
\begin{equation}
(-1)^{\zeta}(-1)^{2\gamma_{n}}=1.\label{126}
\end{equation}
Therefore, taking into account Eq. (\ref{116}), we conclude that
$(-1)^{\zeta}=-1$. In other words, symmetry of the total state vector
under particle exchange, obtained from (\ref{115}) with $(-1)^{\zeta}=-1$,
\begin{equation}
\left|\varPsi\right\rangle =\frac{1}{\sqrt{2}}\left[\left|n\gamma_{n}\right\rangle _{1}\left|m\gamma_{m}\right\rangle _{2}-\left|m\gamma_{m}\right\rangle _{1}\left|n\gamma_{n}\right\rangle _{2}\right].\label{127}
\end{equation}
 implies antisymmetry of the (energy) state vector (\ref{103}),
\begin{equation}
\left|\varPsi\right\rangle =\frac{1}{\sqrt{2}}\left[\left|n\right\rangle _{1}\left|m\right\rangle _{2}-\left|m\right\rangle _{1}\left|n\right\rangle _{2}\right].\label{128}
\end{equation}

\subsection{The Pauli principle}

The above procedure is of course applicable to particles with higher
spin; thus for any half-integer value of $\gamma$, $(-1)^{2\gamma}=-1$
and according to Eq. (\ref{126}) the bipartite (energy) state vector
will be antisymmetric with respect to particle exchange, as in Eq.
(\ref{128}). 

We recall that Eq. (\ref{128}) is valid for $\left|n\right\rangle \neq\left|m\right\rangle $.
If $\left|n\right\rangle =\left|m\right\rangle $ and the spin is
not taken into account, the state vector is simply the product of
the individual energy eigenvectors, $\left|\varPsi\right\rangle =\left|n\right\rangle _{1}\left|n\right\rangle _{2}$;
according to Eq. (\ref{96}) the particle variables are not correlated
and the bipartite system is obviously not entangled. However, with
the introduction of spin, the complete state function is different
from zero for $\left|n\right\rangle =\left|m\right\rangle $, under
the condition that $\left|\gamma_{n}\right\rangle \neq\left|\gamma_{m}\right\rangle $.
If this is the case, Eq. (\ref{127}) is reduced to
\begin{equation}
\left|\varPsi\right\rangle =\frac{\left|n\right\rangle _{1}\left|n\right\rangle _{2}}{\sqrt{2}}\left[\left|\gamma_{1}\right\rangle \left|\gamma_{2}\right\rangle -\left|\gamma_{2}\right\rangle \left|\gamma_{1}\right\rangle \right].\label{130}
\end{equation}
In other words, entanglement can arise from energy degeneracy, if
$\mathcal{\mathcal{E\mathrm{=}}E}_{n}\mathcal{\mathrm{+}E}_{m}$ with
$\mathcal{E}_{n}\mathcal{\mathrm{\neq}E}_{m}$, or from spin degeneracy,
if $\gamma=\gamma_{1}+\gamma_{2}$ with $\gamma_{1}\mathcal{\mathrm{\neq}\gamma}_{2}$.
Since for the electron (and other spin-$1/2$ particles) $\gamma_{i}=\pm\frac{1}{2}$
, Eq. (\ref{130}) takes the form (except for an irrelevant overall
sign)
\begin{equation}
\left|\varPsi\right\rangle =\frac{\left|n\right\rangle _{1}\left|n\right\rangle _{2}}{\sqrt{2}}\left[\left|\frac{1}{2}\right\rangle \left|-\frac{1}{2}\right\rangle -\left|-\frac{1}{2}\right\rangle \left|\frac{1}{2}\right\rangle \right].\label{132}
\end{equation}

In Section \ref{Ent} it was shown that the correlation between particle
variables results from the antiphase response to the single common
field mode of frequency $\omega_{mn}$ with $\mathcal{E}_{n}\mathcal{\mathrm{\neq}E}_{m}$.
On the other hand, when $\left|n\right\rangle =\left|m\right\rangle $,
we noted from Eq. (\ref{66}) that the two particles respond in antiphase
to all (common) field modes; in this case, correlation is established
as a result of the response of both particles to a common field mode
of circular polarization. In other words, the entanglement results
not from the response to a single mode connecting two states separated
by their energies, $\triangle\mathcal{E}_{nm}=\left|\mathcal{E}_{n}\mathcal{-E}_{m}\right|$,
but from a mode connecting two states separated by their spins, $\triangle\gamma_{12}=\left|\gamma_{1}-\gamma_{2}\right|$.
Just as in the first case $\mathcal{\triangle\mathcal{E}\mathrm{=}}\hbar\omega_{mn}$
is the energy exchanged with the field in a transition, in the second
case $\hbar\triangle\gamma_{12}=\hbar$ is the angular momentum exchanged
with the field in a transition.

Equation (\ref{132}) leaves no room for a third electron in the same
energy state $\left|n\right\rangle $ because its spin parameter would
be either equal to $\gamma_{1}$ or to $\mathcal{\gamma}_{2}$. The
conclusion holds for any pair of identical half-integer spins, because
the condition $\triangle\gamma_{ij}=\left|\gamma_{i}-\gamma_{j}\right|=1$
cannot be satisfied simultaneously for $i.j=1,2,3$: if two half-integer
values of $\gamma$ satisfy $\triangle\gamma_{ij}=1$, the third value
of $\gamma$ differs from the first two ones by an even number. To
illustrate, consider $\varGamma_{F}=\frac{3}{2}.$ Possible pairs
$(\gamma_{1},\gamma_{2}$) are $(\frac{3}{2},\frac{1}{2})$, $(\frac{3}{2},\frac{-3}{2})$,
$(-\frac{3}{2},-\frac{1}{2})$; there is no $\gamma_{3}$ that simultaneously
satisfies $\triangle\gamma_{31}=\left|\gamma_{3}-\gamma_{1}\right|=1$
and $\triangle\gamma_{32}=\left|\gamma_{3}-\gamma_{2}\right|=1$. 

This is a clear example of Pauli's exclusion principle. The present
discussion reveals the physical basis of the phenomenon: two particles
in the same energy state respond in antiphase to a single (circularly
polarized) mode of the field and a third particle cannot respond in
antiphase to the first two. 

\section{Discussion }

In this work, the symmetrization postulate and the spin-statistics
theorem were shown to follow from the in-phase or antiphase response
of identical particles to specific modes of the common background
radiation field. The inclusion of spin in the analysis allowed the
identification of the type B and F families introduced in section
\ref{BF} as bosons and fermions, and led to the Pauli exclusion principle
in the case of fermions. 

Key quantum phenomena that were introduced as postulates in the foundational
phase of quantum mechanics, and that have been repeatedly confirmed
both formally and experimentally, thus find a physical justification.
The picture provided by the present approach is very suggestive. In
particular, it shows that the collective behavior of identical particles,
which leads to the respective quantum statistics, is a consequence
of the mediation of specific field modes that \textquotedbl connect\textquotedbl{}
the particles and correlate their dynamics, producing entanglement.
A mysterious, apparently non-local connection between particles, as
described by quantum formalism, is thus shown to be an entirely causal
and local effect of the bridging role of the common background field.
Given the increasing attention paid to entanglement phenomena and
their applications, particularly in the fields of quantum information,
computing and communication, the insight gained from this perspective
should prove highly fruitful.

The results reported in this paper suggest further investigation.
In particular, an extension of the one-dimensional analysis carried
out here to three dimensions would allow an adequate treatment of
more general problems involving additional dynamical variables, including
orbital angular momentum.

\paragraph*{Acknowledgment}

We would like to thank two reviewers for their constructive comments,
which helped to improve the clarity of the exposition.

\paragraph*{Conflict of interest}

The authors declare that the research was conducted in the absence
of any commercial or financial relationships that could be construed
as a potential conflict of interest.

\paragraph*{Author contributions}

Both authors contributed equally to this work.


\begin{thebibliography}{10}
\bibitem{Pauli25} Pauli, W. (1925), Über den Zusammenhang des Abschlusses
der Elektronengruppen im Atom mit der Komplexstruktur der Spektren.
\emph{Zeitschr. f. Physik} 31 (1): 765--783.

\bibitem{Heis26}Heisenberg, W. (1926), Mehrkörperproblem und Resonanz
in der Quantenmechanik. \emph{Zeitschr. f. Physik} 38, 411--426.
11. 

\bibitem{Dirac26}Dirac, P.A.M. (1926), On the theory of quantum mechanics.
\emph{Proc. R. Soc. Lond.} A 112, 661--677.

\bibitem{Kaplan20}Kaplan, I. G. (2020), The Pauli Exclusion Principle
and the Problems of Its Experimental Verification. \emph{Symmetry}
12, 320; doi:10.3390/sym12020320.

\bibitem{Pauli46}Pauli, W. (1946), Exclusion principle and quantum
mechanics. Nobel lecture. https://www.nobelprize.org/uploads/2018/06/pauli-lecture.pdf

\bibitem{Pauli50}Pauli, W. (1950), On the connection between spin
and statistics, \emph{Prog. Theor. Phys}. 5, 526.

\bibitem{TEQ}de la Peña, L., A.M. Cetto and A. Valdés-Hernández (2015),
\emph{The Emerging Quantum. The Physics Behind Quantum Mechanics}.
Springer; doi:10.1007/978-3-319-07893-9. 

\bibitem{FOOP24}Cetto, A. M. and L. de la Peña (2024), The Radiation
Field, at the Origin of the Quantum Canonical Operators. \emph{Found.
Phys.} 54:51; doi:10.1007/s10701-024-00775-5. 

\bibitem{Hunter705}Hunter, G. and I. Schlifer (2005), Explicit Spin
Coordinates. arXiv:quant-ph/0507008v1 1 Ju8l 2008.

\bibitem{Jabs10}Jabs, A. (2010), Connecting spin and statistics in
quantum mechanics. \emph{Found. Phys}. 40, 776. Last revised version
arXiv:quant-ph/0810.2399v4 3 Feb 2014.

\bibitem{Okuda11}Okuda, T. et al (2011), Efficient spin resolved
spectroscopy observation machine at Hiroshima Synchrotron Radiation
Center. \emph{Rev. Sci. Instrum}. 82, 103302; doi:10.1063/1.3648102.

\bibitem{De21}De, A., T. K. Bhowmick and R. K. Lake (2021), Anomalous
magneto-optical effects in an antiferromagnet--topological-insulator
heterostructure. \emph{Phys. Rev. Appl.} 16, 014043.

\bibitem{Sobelman79}Sobelman, I. I. (1979), \emph{Atomic Spectra
and Radiative Transitions}. Springer.

\bibitem{Mandel95}Mandel, L. and E. Wolf (1995), \emph{Optical Coherence
and Quantum Optics}. Casmbridge University Press, Chapter 10.

\bibitem{JPCS14}Cetto, A.M., L. de la Peña, A. Valdés-Hernández (2014),
Emergence of quantization: the spin of the electron. \emph{J. Phys.
Conf. Series} 504, 012007.

\bibitem{JPCS16}Cetto, A.M. and L. de la Peña (2015), Electron system
correlated by the zero-point field: physical explanation for the spin-statistics
connection. \emph{J. Phys. Conf. Series} 701, 012008; doi:10.1088/1742-6596/701/1/012008.

\end{thebibliography}
\end{document}